\documentclass[aps,preprint]{revtex4}%
\usepackage{amssymb}
\usepackage{amsfonts}%
\usepackage{amsmath}%
\usepackage{graphicx}
\providecommand{\U}[1]{\protect\rule{.1in}{.1in}}

\begin{document}
\preprint{HEP/123-qed}
\title[Short title for running header]{Spectroscopy of Rindler Modified Schwarzschild Black Hole}
\author{I. Sakalli}
\affiliation{Department of Physics, Eastern Mediterranean University, Gazimagosa, North
Cyprus, Mersin 10, Turkey}
\author{S.F. Mirekhtiary}
\affiliation{Department of Physics, Eastern Mediterranean University, Gazimagosa, North
Cyprus, Mersin 10, Turkey}

\begin{abstract}
We study quasinormal modes (QNMs) of uncharged Grumiller black hole (GBH).
This massive BH has a Rindler acceleration \$a\$, and hence it is also called
Rindler modified Schwarzschild BH. After reducing the radial equation of the
massless Klein-Gordon equation to the Zerilli equation, we compute the complex
frequencies of the QNMs of the GBH. To this end, an approximation method which
considers small perturbation around its horizon is used. Considering the
highly damped QNMs in the process proposed by Maggiore, the quantum entropy
and area spectra of these BHs are found. Although the QNM frequencies are
tuned by the Rindler term, we show in detail that the spacing does not depend
on it. Here, dimensionless constant $\epsilon$ of the area spectrum is found
to be double of its Schwarzschild value. The latter result is also discussed.

\end{abstract}
\volumeyear{year}
\volumenumber{number}
\issuenumber{number}
\eid{identifier}
\date[Date text]{date}
\received[Received text]{date}

\revised[Revised text]{date}

\accepted[Accepted text]{date}

\published[Published text]{date}

\startpage{1}
\endpage{2}
\maketitle
\tableofcontents

\section{Introduction}

One of the trend subjects in the thermodynamics of BHs is the quantization of
the BH horizon area and entropy. The pioneer works in this regard dates back
to 1970s, in which Bekenstein showed that BH entropy is proportional to the
area of the BH horizon \cite{Bek1,Bek2}. Furthermore, Bekenstein
\cite{Bek3,Bek4,Bek5} conjectured that if the BH horizon area is an adiabatic
invariant, according to Ehrenfest's principle it has a discrete and equally
spaced spectrum as the following%

\begin{equation}
A_{n}=\epsilon nl_{p}^{2},\text{ \ \ \ \ \ \ }(n=0,1,2.......),\label{1}%
\end{equation}

where $\epsilon$ is a dimensionless constant and $l_{p}$ is the Planck length.
$A_{n}$\ denotes the area spectrum of the BH horizon and $n$ is the quantum
number. One can easily see that when the BH absorbs a test particle the
minimum increase of the horizon area is $\Delta A_{\min}=\epsilon l_{p}^{2}$.
In units with $c=G=1$ and $l_{p}^{2}=\hbar,$ the undetermined dimensionless
constant $\epsilon$ is considered as the order of unity. Bekenstein purported
that the BH\ horizon is formed by patches of equal area $\epsilon\hbar$ with
$\epsilon=8\pi$. After that, many studies have been done in order to obtain
such equally spaced area spectrum, whereas the spacing could be different than
$\epsilon=8\pi$ (a reader may refer to~\cite{Jia} and references therein).

Since a\ BH is characterized by mass, charge and angular momentum, it can be
treated as an elementary particle. Since each object made by the elementary
particles has its own characteristic vibrations known as the QNM frequencies,
QNMs should reveal some information about the BH. Specially they are important
for observational aspect of gravitational waves phenomena. In the same
conceptual framework, Hod \cite{Hod1,Hod2} suggested that $\epsilon$\ can be
obtained by using the QNM of a BH. Based on Bohr's correspondence principle
\cite{Bohr}, Hod theorized that the real part of the asymptotic QNM frequency
($\omega_{R}$) of a highly damped BH is associated with the quantum transition
energy between two quantum levels of the BH. This transition frequency allows
a change in the BH mass as $\Delta M=\hbar\omega_{R}$. For the Schwarzschild
BH, Hod calculated the value of the dimensionless constant as $\epsilon=4\ln
3$. Thereafter, Kunstatter \cite{Kunstatter} considered the natural adiabatic
invariant $I_{adb}$ for system with energy $E$ and vibrational frequency
$\Delta\omega$ (for a BH, $E$ is identified with the mass $M$ ) which is given by%

\begin{equation}
I_{adb}=\int\frac{dE}{\Delta\omega}.\label{2}%
\end{equation}

At large quantum numbers, the adiabatic invariant is quantized via the
Bohr-Sommerfeld quantization; $I_{adb}\simeq n\hbar$. Thus, Hod' result
($\epsilon=4\ln3$) is also derived by Kunstatter. Later on, Maggiore
\cite{Maggiore} set up another method in which the QNM of a perturbed BH is
considered as a damped harmonic oscillator. This approach was plausible since
the QNM has an imaginary part. In other words, Maggiore considered the proper
physical frequency of the harmonic oscillator with a damping term in the form
of $\omega=\left(  \omega_{R}^{2}+\omega_{I}^{2}\right)  ^{\frac{1}{2}}$,
where $\omega_{R}$\ and $\omega_{I}$\ are the real and imaginary parts of the
frequency of the QNM, respectively. In the $n\gg1$ limit which is equal to the
case of highly excited mode, $\omega_{I}\gg\omega_{R}$. Therefore, one infers
that $\omega_{I}$\ should be used rather than $\omega_{R}$\ in the adiabatic
quantity. As a result, it was found that $\epsilon=8\pi,$ which corresponds to
the same area spectrum of Bekenstein's original result of the Schwarzschild BH
\cite{Vagenas,Medved}. By this time, we can see numerous studies in the
literature in which Maggiore's method (MM) was employed (see for instance
\cite{Samp1,Samp2,Samp3,Samp4,Samp5,Samp6,Samp7}).

Rindler acceleration "$a$" \cite{Rindler} has recently become popular anew.
This is because of Grumiller's BH (GBH) having a static and spherically
symmetric metric \cite{Grumiller1,Grumiller2} which attempts to describe the
gravity of a central mass at large distances. In \cite{Grumiller1}, it was
suggested that the effective potential of a central gravitating mass $M$
should include $r-$dependent acceleration term. Therefore the problem
effectively degrades to a $2D$ system where the Newton's gravitational force
modifies into $F_{G}=-m(\frac{M}{r^{2}}+a)$ in which $m$ is the mass of a test
particle. For $a<0$, the two forces represent repulsive property however while
$a>0$ gives an inward attractive force. Here, unless stated otherwise,
throughout the paper we shall use $a>0$. Firstly, this attractive acceleration
was hypothesized to decipher the inexplicable acceleration that revealed after
the long period observations on the Pioneer spacecrafts -- Pioneer 10 and
Pioneer 11 -- after they covered a distance about $3\times10^{9}km$ on their
trajectories out of the Solar System \cite{Berman}. The associated
acceleration is unlikely attractive i.e., directed toward the Sun, and this
phenomenon is known as the Pioneer anomaly. However, the newest and widely
acclaimed study \cite{Turyshev} on this subject has shown that the Pioneer
anomaly could be explained by thermal heat loss of the sattelite. Besides
these, it is also speculated that the Rindler acceleration may play the role
of dark matter in galaxies \cite{Grumiller1,Grumiller2}. So, the integration
of the Newton's theory with the Rindler acceleration could explain rotation
curves of spiral galaxies without the presence of a dark matter halo. In
brief, the main role of $a$ is to constitute a rough model involving rotation
curves with a linear growing of the velocity with the radius. Up to the
present, the main studies on the GBH are \cite{Carloni, Sultana} in which
light bending, gravitational redshift and perihelion shifts were computed for
the planets in our solar system. For the most recent work on the GBH, which is
about its geodesics a reader may consult \cite{Halilsoy}. As a last remark,
when the Rindler term in the GBH metric is terminated ($a=0$), all results
reduce to those of Schwrazschild BH as it must.

In this paper, our main motivation is to examine how the influence of the
Rindler acceleration effects the GBH\ spectroscopy.We shall first compute the
QNMs of the GBH and subsequently use them in the MM. For this purpose, we
organize the the paper as follows. In the next section, we describe the GBH
metric and its basic thermodynamical features. We also represent that how the
massless Klein Gordon equation reduces to the one dimensional
Schr\"{o}dinger-type wave equation which is the so-called the Zerilli equation
\cite{Chandra} in the GBH geometry. Sect. 3 is devoted to the calculation of
the QNMs of the GBH by considering the small perturbations around the horizon.
After that, we employ the MM for the GBH in order to compute its entropy and
area spectra. Finally the conclusion is given in Sect. 4.

\section{GBH and its Zerilli equation}

In this section, we represent the geometry and some thermodynamical properties
of the GBH. We explicitly show how one gets the radial equation for a massless
scalar field in the background of the GBH. Then, we give the one-dimensional
Schr\"{o}dinger wave equation form (the so-called Zerilli equation
\cite{Chandra}) of the GBH.

GBH's line element \cite{Grumiller1}\ without cosmological constant is given by%

\begin{equation}
ds^{2}=-H(r)dt^{2}+\frac{dr^{2}}{H(r)}+r^{2}d\Omega^{2},\label{3}%
\end{equation}

where $d\Omega^{2}$ is the standard metric on $2-$sphere and the metric
function $H(r)$ is computed as%

\begin{align}
H(r) &  =1-\frac{2M}{r}+2ar,\nonumber\\
&  =\frac{2a}{r}(r-r_{h})(r-r_{0}),\label{4}%
\end{align}

where $M$ is the mass of the BH and a is a positive real constant, which
corresponds to the Rindler acceleration parameter. One can easily see that
when $a=0$, spacetime (1) reduces to the well-known Schwarzschild BH. On the
other hand, herein $r_{0}$ is found to be%

\begin{equation}
r_{0}=-\frac{\sqrt{1+16aM}+1}{4a},\label{5}%
\end{equation}

which cannot be horizon due to its negative signature. Therefore, the GBH
possesses only one horizon (event horizon, $r_{h}$) which is given by%

\begin{equation}
r_{h}=\frac{\sqrt{1+16aM}-1}{4a},\label{6}%
\end{equation}

such that while $a\rightarrow0$\ we get $r_{h}=2M$, as it is expected. After
computing the scalars of the metric, we obtain%

\begin{align}
K  & =R_{\alpha\beta\mu\nu}R^{\alpha\beta\mu\nu}=32\frac{a^{2}}{r^{2}}%
+48\frac{M^{2}}{r^{6}},\nonumber\\
R  & =-12\frac{a}{r},\label{7}\\
R_{\alpha\beta}R^{\alpha\beta}  & =40\frac{a^{2}}{r^{2}}.\nonumber
\end{align}

which obviously show that central curvature singularity is at $r=0$.

Surface gravity \cite{Wald} of the GBH can simply be calculated through the
following expression%

\begin{equation}
\kappa=\left.  \frac{H^{\prime}(r)}{2}\right\vert _{r=rh}=\frac{a\left(
r_{h}-r_{0}\right)  }{r_{h}},\label{8}%
\end{equation}

where a prime "$\prime$" denotes differentiation with respect to $r$. From
here on in, one obtains the Hawking temperature $T_{H}$ of the GBH as%

\begin{align}
T_{H} &  =\frac{\hbar\kappa}{2\pi},\nonumber\\
&  =\frac{a\left(  r_{h}-r_{0}\right)  \hbar}{2\pi r_{h}},\label{9}%
\end{align}

From the above expression, it is seen that while the GBH losing its $M$ by
virtue of the Hawking radiation, $T_{H}$ increases (i.e., $T_{H}%
\rightarrow\infty$) with $M\rightarrow0$ in such a way that its divergence
speed is tuned by $a$. Meanwhile, one can check that $\lim_{a\rightarrow
0}T_{H}=\frac{1}{8\pi M}$ which is well-known Hawking temperature computed for
the Schwarzschild BH. The Bekenstein-Hawking entropy is given by%

\begin{align}
S_{BH} &  =\frac{A_{h}}{4\hbar},\nonumber\\
&  =\frac{\pi r_{h}^{2}}{\hbar},\label{10}%
\end{align}

Its differential form is written as%

\begin{equation}
dS_{BH}=\frac{4\pi}{\sqrt{1+16aM}\hbar}r_{h}dM,\label{11}%
\end{equation}

By using the above equation, the validity of the first law of thermodynamics
for the GBH can be approved via%

\begin{equation}
T_{H}dS_{BH}=dM.\label{12}%
\end{equation}

In order to find the entropy spectrum by using the MM, here we shall \ firstly
consider the massless scalar wave equation on the geometry of the GBH. The
general equation of massless scalar field in a curved spacetime is written as%

\begin{equation}
\square\Psi=0,\label{13}%
\end{equation}

where $\square$ denotes the Laplace-Beltrami operator. Thus, the above
equation is equal to%

\begin{equation}
\frac{1}{\sqrt{-g}}\partial_{i}(\sqrt{-g}\partial^{i}\Psi),\text{
\ \ \ }i=0...3,\label{14}%
\end{equation}

Using the following ansatz for the scalar field $\Psi$ in the above equation%

\begin{equation}
\Psi=\frac{R(r)}{r}e^{i\omega t}Y_{L}^{m}(\theta,\varphi),\text{
\ \ }Re(\omega)>0,\label{15}%
\end{equation}

in which $Y_{L}^{m}(\theta,\varphi)$ is the well-known spheroidal harmonics
which admits the eigenvalue $-L(L+1)$ \cite{Du}, one obtains the following
Zerilli equation \cite{Chandra} as%

\begin{equation}
\left[  -\frac{d^{2}}{dr^{\ast2}}+V(r)\right]  R(r)=\omega^{2}R(r),\label{16}%
\end{equation}

where the effective potential is computed as%

\begin{equation}
V(r)=H(r)\left[  \frac{L(L+1)}{r^{2}}+\frac{2M}{r^{3}}+\frac{a}{r}\right]
,\label{17}%
\end{equation}

The tortoise coordinate $r^{\ast}$ is defined as,%

\begin{equation}
r^{\ast}=\int\frac{dr}{H(r)},\label{18}%
\end{equation}

which yields%

\begin{equation}
r^{\ast}=\frac{1}{2a(r_{h}-r_{0})}\ln\left[  \frac{(\frac{r}{r_{h}}-1)^{rh}%
}{(r-r_{_{0}})^{r_{0}}}\right]  ,\label{19}%
\end{equation}

Finally, one can easily check the asymptotic limits of $r^{\ast}$ as follows%

\begin{equation}
\lim_{r\rightarrow rh}r^{\ast}=-\infty\text{ \ and }\lim_{r\rightarrow\infty
}r^{\ast}=\infty.\label{20}%
\end{equation}

\section{QNMs and entropy/area spectra of GBH}

In this section, we intend to derive the entropy and area spectra of the GBH
by using the MM. Gaining inspiration from the studies
\cite{Appro1,Appro2,Appro3}, here we use an approximation method in order to
define the QNMs. According to this method, to obtain QNM it is sufficient to
use one of the two definitions of the QNMs; only ingoing waves should exist
near the horizon. Namely, %

\begin{equation}
\left.  R(r)\right\vert _{QNM}\sim e^{i\omega r^{\ast}}\text{ at }r^{\ast
}\rightarrow-\infty,\label{21}%
\end{equation}

Now we can proceed to solve Eq. (16) in the near horizon limit and then impose
the above boundary condition to find the frequency of QNM i.e., $\omega$.
Expansion of the metric function $H(r)$ around the event horizon is given by%

\begin{align}
H(r) &  =H^{\prime}(r_{h})(r-r_{h})+\Game(r-r_{h})^{2},\nonumber\\
&  \simeq2\kappa(r-r_{h}),\label{22}%
\end{align}

where $\kappa$\ is the surface gravity, which is nothing but $\frac{1}%
{2}H^{\prime}(r_{h})$. From Eq. (18) we now obtain%

\begin{equation}
r^{\ast}\simeq\frac{1}{2\kappa}\ln(r-r_{h}),\label{23}%
\end{equation}

Furthermore, after setting $y=r-r_{h}$ and inserting Eq. (22) into Eq. (17)
together with performing Taylor expansion around $r=r_{h}$, we find the near
horizon form of the effective potential as%

\begin{equation}
V(y)\simeq2\kappa y\left[  \frac{L(L+1)}{r_{h}^{2}}(1-\frac{2y}{r_{h}}%
)+\frac{2\kappa}{r_{h}}(1-\frac{y}{r_{h}})\right]  .\label{24}%
\end{equation}

After substituting Eq. (24) into the Zerilli equation (16), one gets%

\begin{equation}
-4\kappa^{2}y^{2}\frac{d^{2}R(y)}{dy^{2}}-4\kappa^{2}y\frac{dR(y)}%
{dy}+V(y)R(y)=\omega^{2}R(y),\label{25}%
\end{equation}

Solution of the above equation yields%

\begin{equation}
R(y)\sim y_{1}^{\frac{i\omega}{2\kappa}}{}_{1}F_{1}(\widehat{a},\widehat{b}%
;\widehat{c}),\label{26}%
\end{equation}

where $_{1}F_{1}(\widehat{a},\widehat{b};\widehat{c})$\ is the confluent
hypergeometric function \cite{Abramowitz}. The parameters of the confluent
hypergeometric functions are found to be%

\begin{align}
\widehat{a} &  =\frac{1}{2}+i(\frac{\omega}{2\kappa}-\frac{\hat{\alpha}}%
{\hat{\beta}\sqrt{\kappa}}),\nonumber\\
\widehat{b} &  =1+i\frac{\omega}{\kappa},\label{27}\\
\widehat{c} &  =i\frac{\hat{\beta}x}{2r_{h}\sqrt{\kappa}},\nonumber
\end{align}

where%

\begin{align}
\hat{\beta} &  =4\sqrt{r_{h}}\sqrt{L(L+1)+\kappa r_{h}},\nonumber\\
\hat{\alpha} &  =L(L+1)+2\kappa r_{h},\label{28}%
\end{align}

In the limit of $y\ll1$, the solution (26) becomes%

\begin{equation}
R(y)\sim C_{1}y^{-\frac{i\omega}{2\kappa}}\frac{\Gamma(i\frac{\omega}{\kappa
})}{\Gamma(\widehat{a})}+C_{2}y^{\frac{i\omega}{2\kappa}}\frac{\Gamma
(-i\frac{\omega}{\kappa})}{\Gamma(1+\widehat{a}-\widehat{b})},\label{29}%
\end{equation}

where constants $C_{1}$ and $C_{2}$ represent the amplitudes of the
near-horizon outgoing and ingoing waves, respectively. Now, since there is no
outgoing wave in the QNM at the horizon, the first term of Eq. (29) should be
terminated. This is possible with the poles of the Gamma function of the
denominator. Therefore, the poles of the Gamma function are the policy makers
of the frequencies of the QNMs. Thus, the frequencies of the QNM of the GBH
are read as%

\begin{equation}
\omega_{s}=\frac{2\sqrt{\kappa}\hat{\alpha}}{\hat{\beta}}+i\frac{2\pi}{\hbar
}(2s+1)T_{H},\text{ \ \ \ \ \ \ }(s=1,2,3,...)\label{30}%
\end{equation}

where $m$\ is the overtone quantum number of the QNM. Thus, the imaginary part
of the frequency of the QNM is%

\begin{equation}
\omega_{I}=\frac{2\pi}{\hbar}(2s+1)T_{H},\label{31}%
\end{equation}

As it can be seen from above, the Rindler acceleration plays a crucial role on
$\omega_{I}$. While $a\rightarrow0$ , $\omega_{I}=\frac{(2s+1)}{4M}$ which is
consistent with the Schwarzschild BH result \cite{Chen,Sakalli}. Hence the
transition frequency between two highly damped neighboring states becomes
$\Delta\omega\equiv\Delta\omega_{I}=\omega_{s+1}-\omega_{s}=4\pi T_{H}/\hbar$.
Hence, the adiabatic invariant quantity (2) turns out to be%

\begin{equation}
I_{adb}=\frac{\hbar}{4\pi}\int\frac{dM}{T_{H}},\label{32}%
\end{equation}

According to the first law of thermodynamics (12), it reads%

\begin{equation}
I_{adb}=\frac{S_{BH}}{4\pi}\hbar,\label{33}%
\end{equation}

Finally, recalling the Bohr-Sommerfeld quantization rule $I_{adb}=\hbar n,$
one gets the spacing of the entropy spectrum as%

\begin{equation}
S_{n}=4\pi n,\label{34}%
\end{equation}

Since $S=\frac{A}{4\hbar},$ the area spectrum is obtained as%

\begin{equation}
A_{n}=16\pi n\hbar,\label{35}%
\end{equation}

From the above, we can simply measure the area spacing as%

\begin{equation}
\Delta A=16\pi\hbar.\label{36}%
\end{equation}

It is easily seen that unlike to $\omega_{I}$ the spectroscopy of the GBH is
completely independent of the Rindler term $a$. The obtained spacings between
the levels are double of the Bekenstein's original result which means that
$\epsilon=16\pi$. The discussion on this discrepancy is made in the conclusion part.

\section{Conclusion}

In this paper, the BH spectroscopy of the GBH is investigated through the MM.
We applied an approximation method given in \cite{Appro1,Appro2,Appro3} to the
Zerilli equation (16) in order to compute the QNM of the GBH. After a
straightforward calculation, the QNM frequency of the GBH are analytically
found. The obtained result shows that imaginary part of the frequency,
$\omega_{I}$ depends on the Rindler term $a$. Then, with the aid of Eq. (2),
we obtained the entropy/area spectra of the GBH. Both spectra are found to be
independent of the Rindler term $a,$ and they are equally spaced. Futhermore,
we read the dimensionless constant as $\epsilon=16\pi$ which means that the
equi-spacing is the double of its Schwarzschild value: $\epsilon=8\pi$
\cite{Maggiore}. This differentness may arise due to the Schwinger mechanism
\cite{SKim}. Because, in the Bekenstein's original work \cite{Bek2}, one gets
the entropy spectrum by combining both the Schwinger mechanism and the
Heisenberg quantum uncertainty principle. However, the method that we applied
here considers only the uncertainty principle via the Bohr-Sommerfeld
quantization. Therefore, as stated in \cite{Hod2}, the spacings between two
neighboring levels may become different depending on the which method is
preferred. Thus, finding $\epsilon=16\pi$ rather than its known value
$\epsilon=8\pi$ is not an unexpected result. As a last remark, evenly spaced
structure of the entropy and area spectra of the GBH is also in agreement with
the Wei et al.'s conjecture \cite{Samp3}, which proposes that static BHs of
Einstein's gravity theory has equidistant quantum spectra of both entropy and
area .

\end{document}